\documentclass[]{epl2}
\usepackage{graphicx}

\begin{document}

\newcommand{\BiTe} {Bi$_2$Te$_3$}

\title{Spin-orbit interaction effect in the electronic structure of \BiTe \ observed by angle-resolved photoemission spectroscopy}
\shorttitle{SOI effect in the electronic structure of \BiTe \ observed by ARPES}

\author{
Han-Jin Noh\inst{1,2}\thanks{E-mail: \email{ffnhj@chonnam.ac.kr}} \and H. Koh\inst{2} \and S.-J. Oh\inst{2} \and J.-H. Park\inst{3} \and H.-D. Kim\inst{4} \and J. Rameau\inst{5,6} \and T. Valla\inst{5} \and T. Kidd\inst{5} \and P. D. Johnson\inst{5} \and Y. Hu\inst{5} \and Q. Li\inst{5} } \shortauthor{H.-J. Noh \etal}

\institute{ \inst{1} Department of Physics, Chonnam National University - Gwangju 500-757, Korea \\
\inst{2} School of Physics \& Center for Strongly Correlated Materials Research, Seoul National University - Seoul 151-742, Korea \\
\inst{3} Department of Physics, Pohang University of Science and Technology - Pohang 790-784, Korea  \\
\inst{4} Pohang Accelerator Laboratory, Pohang University of Science and Technology - Pohang 790-784, Korea  \\
\inst{5} Condensed Matter Physics and Materials Science Department, Brookhaven National Laboratory, Upton, New York 11973, USA \\
\inst{6} Department of Physics \& Astronomy, Stony Brook University, Stony Brook, New York 11974, USA \\
}

\date{\today}


\abstract{
The electronic structure of $p$-type doped \BiTe \ is studied by angle resolved photoemission spectroscopy (ARPES) to experimentally confirm the mechanism responsible for the high thermoelectric figure of merit.
Our ARPES study shows that the band edges are located off the $\Gamma$-Z line in the Brillouin zone, which provides direct observation that the spin-orbit interaction is a key factor to understand the electronic structure and the corresponding thermoelectric properties of \BiTe.
Successive time dependent ARPES measurement also reveals that the electron-like bands crossing E$_F$ near the $\underline{\Gamma}$ point are formed in an hour after cleaving the crystals.
We interpret these as surface states induced by surface band bending, possibly due to quintuple inter-layer distance change of \BiTe.
}

\maketitle


It is of great interest to find a good thermoelectric (TE) material competitive to conventional compressor-based active cooling/power generation systems.
Desirable features such as lack of noise or emission of polluting gases, space saving, high reliability and semi-permanent life make them very attractive though to date their main shortcoming has been their low efficiency \cite{DiSalvo1}.
Appropriately doped \BiTe \ has been one of the most competitive candidate materials for fulfilling the above purpose since its high efficiency was discovered 70 years ago \cite{Lange1}.
Since that time, much effort has been expended to improve its TE properties by doping and alloying, but little progress was made until the reports for Bi$_{2-x}$Sb$_x$Te$_{3-y}$Se$_y$ alloys \cite{Stordeur}.
Even though higher efficiency at room temperature was reported in \BiTe/Sb$_2$Te$_3$ superlattice \cite{Venkata1}, it is very important to completely understand the electronic structure of \BiTe \ in order to establish a reliable guideline for finding or synthesizing better TE compounds.

Narrow gap semiconductors composed of covalently bonded heavy elements have generally been considered as prominent TE materials because the gap and covalent bonding (heavy elements) have a tendency to increase (decrease) the electron entropy and mobility (the thermal conductivity), respectively \cite{Mahan1}.
In addition to these general tendencies, a few system-dependent features have also been pointed out for \BiTe; highly anisotropic (pseudo) two-dimensional structure \cite{Hicks1}; the six-valley electron (hole) pockets in $n$($p$)-type doped \BiTe \ \cite{Drabble, Austin, Mallinson, Kohler}.
Among these features, the six-valley electron and hole pockets in the electronic structure of \BiTe \ drew considerable attention.
Since the multivalley conduction band structure of \BiTe \ was inferred from magnetoresistance and Faraday rotation experiments \cite{Drabble, Austin}, the positions and angles of the hole and electron pockets have been investigated by de Haas-van Alphen and Shubnikov-de Haas experiments \cite{Mallinson, Kohler}.
However, a controversy about the six-valley model arose in theoretical studies.
The multi-valley structure had not been perfectly reproduced by the band calculations at an early stage \cite{Katsuki, Oleshko, Pecheur}.
This controversy was not settled until spin-orbit interactions were included in the band calculation scheme.
Even in this case, not all the methods including spin-orbit interaction give the same results.
Early calculations gave six valence band maxima (VBM) and two conduction band minima (CBM) \cite{Thomas, Mishra, Larson1}, but recent \emph{ab initio} calculation studies give six extrema for both the valence bands and the conduction bands, showing that the spin-orbit interaction is a key factor for understanding the electronic structure and the TE properties of \BiTe \ \cite{Youn, Kim, Scheidemantel, Wang}.
However, the issue is not clearly resolved yet.
Although the six-valley model has been proposed for forty years and many theoretical works have been undertaken to explain it, the model itself is based on only indirect experimental facts; no direct evidence has not been provided so far.
Establishing direct evidence is still crucial for confirming whether the spin-orbit interaction is really a dominant factor in determining the electronic structure of Bi 6$p$ orbitals hybridized with Te 5$p$ orbitals.

Angle-resolved photoemission spectroscopy (ARPES) is the most direct method for investigating the electronic band structure of solids.
Photoemission process in a solid corresponds to single-particle excitation from an occupied state to unoccupied one, so the photoelectron contains information of the energy and the momentum inside the solid, i.e. the electronic structure of the solid.
Owing to this direct connection between the observables and the objectives, ARPES plays an important role in researching hot-debated materials.
To date, we are aware of only one ARPES experiment on \BiTe \ that has been reported in the literature \cite{Greanya}.
The angular resolution of the spectra from that experiment was insufficient to address the issue cited above.
In this paper, we present a direct observation of the six-valley model in $p$-type doped \BiTe \ by ARPES.
Our ARPES measurements with a high angular resolution reveal that the valence band maxima (VBM) are located off the $\Gamma$-Z line in the Brillouin zone (BZ) as predicted in the {\it ab initio} band calculation including spin-orbit interaction \cite{Youn, Kim, Scheidemantel, Wang}.
Further, we find that surface states are formed near $\underline{\Gamma}$ point in an hour after cleaving the sample \emph{in situ}, which has obstructed direct observation of the correct bulk electronic structure of \BiTe.

\begin{figure}
\includegraphics[width = 8.6 cm]{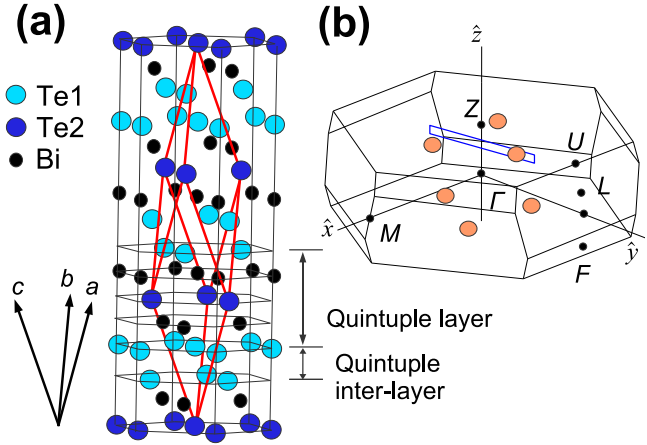}
\caption{\label{BZ}(Color on-line) (a) Crystal structure of \BiTe.
The red rhombohedron represents a rhombohedral unit cell.
(b) Rhombohedral Brillouin zone of \BiTe.
The scale of the drawing is exaggerated along $\Gamma$Z for the purpose of clarity.
The six orange spheres represent the hole pockets in the six valley model.
The small blue segment containing one hole pocket represents the measured $k$ position of the valence band ARPES spectra.
}
\end{figure}

Two kinds of \BiTe \ single crystals, Bi 1\% excess (Bi-1\%) and Te 4\% excess (Te-4\%), were grown by melting Bi and Te in a mixture with the molar ratio and slowly cooling.
Resistivity measurements using the standard four probe method show both of samples (Bi-1\% and Te-4\%) are metallic.
Seebeck coefficient and Hall coefficient measurements show the samples are $p$-type.
For ARPES experiments, the samples were attached to a cryostat and introduced into our analysis chamber in ultra-high vacuum environment of 5$\times$10$^{-11}$ Torr.
A shiny cleaved surface was obtained by the top post method {\it in situ}, and the sample orientation was checked by low energy electron diffraction (LEED).
Low photon energy measurements ($\hbar\omega \sim$20 eV) were performed on undulator beam line U13UB at the National Synchrotron Light Source with a Scienta SES-2002 electron spectrometer.
High photon energy measurements ($\hbar\omega \sim$121 eV) were done on undulator beam line U10 at the Pohang Light Source.
The instrumental energy resolution and momentum resolution were set to 20 meV (30 meV) and $\pm 0.1^{\circ}$ for the low (high) photon energy, respectively.
The reported spectra here, if not otherwise specified, were obtained at 20 K within 15 minutes of cleaving the samples.

The crystal structure of \BiTe \ is rhombohedral with the space group $D^{5}_{3d}$ ($R\bar{3}m$).
The corresponding Brilloune zone (BZ) with high symmetry points is depicted in Fig.~\ref{BZ}.
Along the rhombohedral [111] direction, so-called quintuple layers (Te(1)-Bi-Te(2)-Bi-Te(1)) are stacked, forming intralayer Bi 6$p$-Te 5$p$ covalent bonding and interlayer van der Waals bonding.
The measured $k$ positions in the BZ are depicted as a blue segment in Fig.~\ref{BZ}(b).
In principle, the $k_z$ value can be determined if we use photon energy dependence of the photoemission spectra.
Unfortunately, due to the surface effects, discussed below, the band dispersion along the $k_z$ direction is very weak.
Because of this attempts at determining the position of $k_z$ by changing the photon energy were not successful.
Instead, we use the constant inner potential approximation with the inner potential, $V_0$= 10 eV, to determine the value of $k_z$ \cite{Greanya, Chiang}.
The normal emission spectra taken with $\hbar\omega$=24.0 eV then approximately corresponds to the Z point in the BZ, and so we chose $\hbar\omega$=23.0 eV to search for the valence band maxima in this study.

\begin{figure}
\includegraphics[width = 8.6 cm]{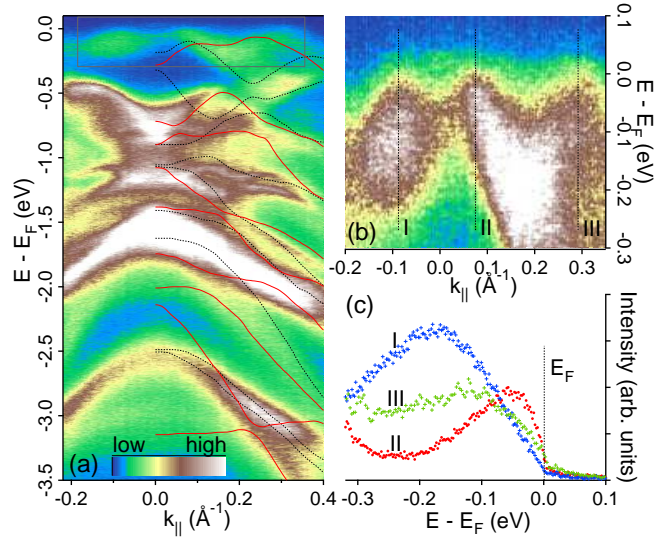}
\caption{\label{VB}(Color on-line)
(a) ARPES intensity map obtained at 20 K with $\hbar\omega$=23.0 eV photon source for the valence bands of Te-4\% along the $\Gamma$ZL mirror plane.
The black dotted (red solid) lines are the band structure calculated by LDA method along the $\Gamma$-L (Z-F) line\cite{Youn}.
The grey box in the top part of the ARPES image is seen at length in (b).
The intensity of the ARPES image is drawn in terrain color scale as the scale bar shows.
(b) Fine view of the ARPES intensity map for the valence band edge.
The right two band edges cross the Fermi level while the left one does not.
This is more clearly seen in the energy distribution curves (EDC) along line I, II, and III.
(c) EDC's along line I, II, and III. The Fermi edge is seen in EDC II and III.}
\end{figure}

ARPES images of the valence bands of Te-4\% are shown in Fig.~\ref{VB}(a).
The valence band structure of Bi-1\% (not shown) was measured and found to be very similar to that of Te-4\%.
For comparison, the bands calculated by the local density approximation (LDA) method along the $\Gamma$-L line and Z-F lines are drawn over the spectra in black dotted lines and red solid lines, respectively \cite{Youn}.
Comparison of the ARPES spectra to both sets of calculations shows that the measured bands do not match either scenario well along lines of high symmetry.
Interestingly, the shallow bands above -1.0 eV more resemble to those of the Z-F line and the deep bands resemble those of the $\Gamma$-L line more closely.

More useful information about the six valley model is contained in the region of the grey rectangular box in Fig.~\ref{VB}(a), which is seen at length in Fig.~\ref{VB}(b).
Here, three hole pocket-like structures are seen as designated with the black dotted lines, I, II, and III, respectively.
In order to check whether those structures are really hole pockets, the energy distribution curves along the lines, I, II, and III, are plotted in Fig.~\ref{VB}(c).
Interestingly, band II and III cross the Fermi level, which means they form hole pockets, while band I does not as shown in Fig.~\ref{VB}(c).
These hole pockets, II and III, need comparing with the theoretical prediction in detail \cite{Youn}.
According to their LDA result, $p$-type doped \BiTe \ has the VBM at (0.546, 0.383, 0.383) and the second highest band edge at (0.665, 0.586, 0.586) in rhombohedral lattice coordinate system.
The corresponding points of the VBM and the second VBM are located at 0.27 \AA$^{-1}$ and 0.13 \AA$^{-1}$ off the $\Gamma$-Z line towards the Z-U direction with $z$=0.88Z and $z$=1.2Z, respectively.
The positions of the VBM are depicted as the orange spheres in Fig.~\ref{BZ}(b).
These positions are directly compared to our experimental results.
As is described above, the two hole pockets, III and II, correspond to the VBM and the second VBM, and each $k_{||}$ value, $\sim$0.29 \AA$^{-1}$ and $\sim$0.09 \AA$^{-1}$, respectively, match the theoretical values well.
Because \BiTe \ has three-fold rotational symmetry around $\Gamma$-Z axis and two-fold rotational symmetry around $\Gamma$-M axis, the system has six VBM's and six second VBM's in the BZ.
This is direct evidence in support of the six-valley model of $p$-type doped \BiTe.
The only ambiguity about the six-valley model is which set is the lowest in energy.
Unfortunately, experimentally answering this question is not trivial because the predicted energy difference between them is only 3.8 meV \cite{Youn}, and the time available for measurement of the bulk states is very limited.
This last constraint is discussed at length below.
On theoretical aspects also, there is a controversy.
Recent LDA calculations have given somewhat different results to the question \cite{Kim, Wang}.
In this respect, our measurements have not been successful at proving a definitive model.

Having established the existence of the multi-valley hole pockets and confirmed the effect of the spin-orbit interaction (SOI) on the electronic structure of \BiTe, it is desirable to understand how the multiple hole pockets induced by the SOI effect enhance the TE properties.
Elements of high atomic number like bismuth have large SOI because the spin-orbit coupling constant is proportional to the atomic number.
As Mahan showed, the TE figure of merit, $ZT$, is a monotonically increasing function of the B parameter:
$B_j = Te\mu n_j(T)(k_B/e)^2/(4\kappa_L)$,
$n_j(T) = N_j(2\pi m_j k_B T/h^2)^{3/2}$,
where $e$ is the carrier charge, $\mu$ is the mobility, $\kappa_L$ is the lattice thermal conductivity, $m_j$ is the density of states mass of the $j$-th band, and $N_j$ is the degeneracy of the band.
Increasing $B_j$ is therefore important for increasing $ZT$ \cite{Mahan}.
Obviously, one of the most efficient ways to increase B is to increase the degeneracy, $N_j$, as in the case of \BiTe.
The above reasoning and our experimental results show that the SOI effect is crucial to understanding the electronic structure of $p$-type doped \BiTe \ with the multiple hole pockets in the BZ, and is an important factor in explaining the high TE figure of merit of \BiTe.

\begin{figure}
\includegraphics[width = 8.6 cm]{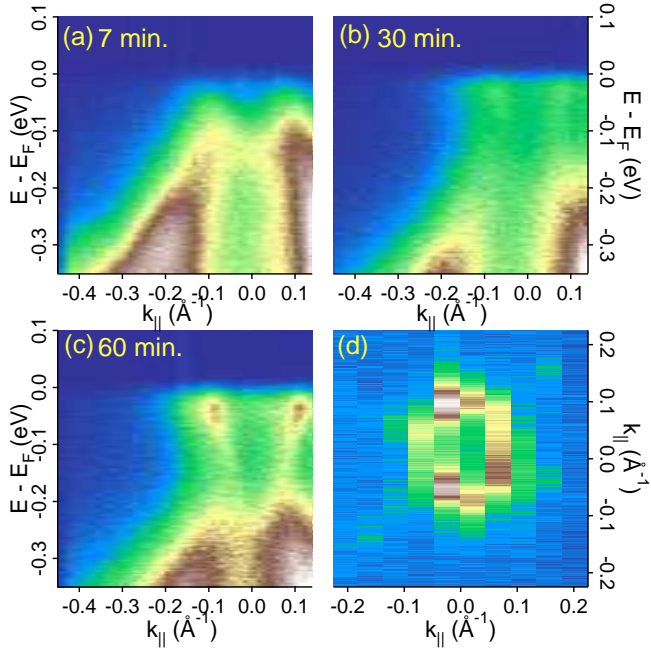}
\caption{\label{SB}(Color on-line)(a) Time dependence of the ARPES spectra after cleaving. The elapsed time is (a) 7 minutes, (b) 30 minutes, and (c) 60 minutes.
(d) Constant energy map of the surface state at E=E$_F$. The spectral weight is integrated over an energy range of E$_F \pm 5$ meV.
}
\end{figure}

As a phenomenon directly impacting ARPES measurements of \BiTe, we also report on surface states forming at a cleaved surface of \BiTe \ crystals.
In most cases of our ARPES experiments, we observed one or two electron pocket-like structures on the $p$-type doped \BiTe \ crystals, both Bi-1\% and Te-4\% at any temperature ranging from 10 to 300 K.
If these states represent the bulk properties of \BiTe \ their existence definitely contradicts our transport measurements.
Fortunately, the time-dependent ARPES study gives very useful information about the origin of the states.
Figure~\ref{SB}(a)-\ref{SB}(c) show the variation of the valence bands of Te-4\% in the ARPES spectra with the elapsed time after cleaving the samples.
When the elapsed time is 7 minutes, it is clearly seen that the valence band edge, located off the $\Gamma$Z line, almost reaches to E$_F$.
As time passes, the valence band splits into two bands.
The valence bands get lowered in energy, and a new band appears crossing E$_F$ in about an hour.
The Fermi surface of the newly emerging state is circular as shown in Fig.~\ref{SB}(d).
This elapsed time dependence suggests that these bands are not the normal surface states induced by translational symmetry breaking owing to cleavage of the bulk crystal.
Otherwise, they would be seen immediately after cleaving the sample.
The endurance test of surface contamination also supports this conjecture.
If the surface electron pocket is formed by impurities such as excess Bi or Te atoms the impurity concentration estimated from the pocket size would be $\sim$1.5\%, which is detectable by high resolution x-ray photoemission spectroscopy (XPS).
In order to check this possibility, we have also performed the shallow core level XPS experiments, but did not observe any noticeable change in the Bi 5$d$ and Te 4$d$ XPS spectra with increase of the elapsed time.
The intensity ratio change of Bi 5$d$ to Te 4$d$ peaks was not observed either.
This indicates that the states are not relevant to the elements in a different chemical environment such as the intercalated excess Bi or Te atoms.

As to the origin of the surface bands, the surface electronic structure of \BiTe \ calculated by the density functional theory provides relevant information (see Fig.~5 of Ref.\cite{Urazhdin1}).
The report points out that the surface band structure of \BiTe \ is different from the bulk one due to the weak inter-layer bonding of the layered compound.
The main differences are the lowering of conduction bands almost to the VBM point, the disappearance of the six-fold degeneracy of the conduction band minima (CBM), and non-dispersive behavior of the bands along the $k_z$ direction.
Most features of our ARPES data are consistent with the argument of surface effects induced by the weak inter-layer bonding, if we assume the electronic structure change in Fig.~\ref{SB}(a)-\ref{SB}(c) is due to slowly relaxing of the quintuple inter-layer distance.
This assumption is not unreasonable when we consider the plasticity of \BiTe.
The main reason for the CBM lowering its energy in the surface electronic structure is the weakening of the covalent $pp\sigma$-bonding along the chain direction of \BiTe, i. e. Te(I)-Bi-Te(II)-Bi-Te(I), induced by an increase of the quintuple interlayer distance.
Because the CBM is an antibonding state, bond weakening causes the state to be lowered in energy.
Inevitably, this lifts up the surface Fermi level and induces surface band bending.

Finally, it is worth mentioning that our finding may require re-interpretation of the previously reported photoemission spectra \cite{Greanya}.
Our experimental data shows that the true bulk electronic structure of \BiTe \ can be obtained only within $\sim$15 minutes after cleaving the samples.
If the sample ages more than this time, the structure near the Fermi level actually consists of surface states.
Further, these states are always seen whenever a good cleave was obtained \emph{in situ} regardless of photon exposure and sample temperatures.
The energy distribution curve from the ARPES image of the surface states shows very similar structure to the conduction bands or the impurity states near the Fermi level in the previous data.

In summary, the electronic structure of $p$-type doped \BiTe \ has been studied by high resolution ARPES.
The valence band edges are located off the $\Gamma$-Z line in the BZ, which directly confirms that the spin-orbit interaction is a dominant factor in understanding the electronic structure and the corresponding thermoelectric properties of \BiTe.
Successive time dependent ARPES measurements also reveal that the electron-like bands crossing the Fermi level near the $\underline{\Gamma}$ point are formed within an hour after cleaving single \BiTe \ crystals.
These are interpreted as surface states induced by surface band bending possibly due to quintuple inter-layer distance change of \BiTe.

\begin{acknowledgments}
This work is supported by KOSEF through CSCMR at SNU and eSSC at POSTECH.
H.-J.N. is supported by the post-BK21 program at CNU.
The work at BNL is supported by DOE.
\end{acknowledgments}


\begin{thebibliography}{}
\bibitem{DiSalvo1} \Name{DiSalvo Francis J.} \REVIEW{Science}{285}{1999}{703}.
\bibitem{Lange1} \Name{Lange P. W.} \REVIEW{Naturwissenschaften}{27}{1939}{133}.
\bibitem{Stordeur} \Name{Stordeur M.} \REVIEW{Phys. Status Solidi}{161}{1990}{831}; \Name{Jeon H.-W., Ha H.-P., Hyun D.-B., and Shim J.-D.} \REVIEW{J. Phys. Chem. Solids}{52}{1991}{579}.
\bibitem{Venkata1} \Name{Venkatasubramanian R., Sillvola E., Colpitts T., and O'Quinn B.} \REVIEW{Nature}{413}{2001}{597}.
\bibitem{Mahan1} \Name{Mahan G., Sales B., and Sharp J.} \REVIEW{Physics Today}{March}{1997}{42}.
\bibitem{Hicks1} \Name{Hicks L. D. and Dresslhaus M. S.} \REVIEW{Phys. Rev. B}{47}{1993}{12727}.
\bibitem{Drabble} \Name{Drabble J. R., Groves R. D., and Wolfe R.} \REVIEW{Proc. Phys. Soc.}{71}{1958}{430}.
\bibitem{Austin} \Name{Austin J. R.} \REVIEW{Proc. Phys. Soc.}{76}{1960}{169}.
\bibitem{Mallinson} \Name{Mallinson R. B., Rayne J. A., and Ure R. W. Jr.} \REVIEW{Phys. Rev.}{175}{1968}{1049}.
\bibitem{Kohler} \Name{K\"ohler H.} \REVIEW{Phys. Stat. Sol. (b)}{73}{1976}{95}; \REVIEW{ibid}{74}{1976}{591}.
\bibitem{Katsuki} \Name{Katsuki S.} \REVIEW{J. Phys. Soc. Jpn.}{26}{1969}{58}.
\bibitem{Oleshko} \Name{Oleshko E. V. and Korolyshin V. N.} \REVIEW{Sov. Phys. Solid State} {27}{1985}{1723}.
\bibitem{Pecheur} \Name{Pecheur P. and Toussaint G.} \REVIEW{Phys. Lett. A} {135}{1989}{223}.
\bibitem{Thomas} \Name{Thomas G. A., Rapkine D. H., Van Dover R. B., Matteiss L. F., Sunder W. A., Schneemeyer L. F., and Waszczak J. V.} \REVIEW{Phys. Rev. B} {46} {1992} {1553}.
\bibitem{Mishra} \Name{Mishra S. K., Satpathy S., and Jepsen O.} \REVIEW{J. Phys.:Condens. Matter} {9} {1997} {461}.
\bibitem{Larson1} \Name{Larson P., Mahanti S. D., and Kanatzidis M. G.} \REVIEW{Phys. Rev. B} {61} {2000} {8162}.
\bibitem{Youn} \Name{Youn S. J. and Freeman A. J.} \REVIEW{Phys. Rev. B} {63} {2001} {085112}.
\bibitem{Kim} \Name{Kim M., Freeman A. J., and Geller C. B.} \REVIEW{Phys. Rev. B} {72} {2005} {035205}.
\bibitem{Scheidemantel} \Name{Scheidemantel T. J., Ambrosch-Draxl C., Thonhauser T., Badding J. V., and Sofo J. O.} \REVIEW{Phys. Rev. B} {68} {2003} {125210}.
\bibitem{Wang} \Name{Wang G. and Cagin T.} \REVIEW{Phys. Rev. B} {76} {2007} {075201}.
\bibitem{Greanya} \Name{Greanya V. A., Tonjes W. C., Liu Rong, Olson C. G., Chung D.-Y., and Kanatzides M. G.} \REVIEW{Phys. Rev. B} {62} {2000} {16425}.
\bibitem{Chiang} \Name{Chiang T. C., Knapp J. A., Aono M., and Eastman E. E.} \REVIEW{Phys. Rev. B} {21} {1980} {3513}.
\bibitem{Mahan} \Name{Mahan G. D.} \REVIEW{J. Appl. Phys.} {65} {1989} {1578}.
\bibitem{Urazhdin1} \Name{Urazhdin S., Bilc D., Mahanti S. D., Tessmer S. H., Kyratsi T., and  Kanatzidis M. G.} \REVIEW{Phys. Rev. B} {69} {2004} {085313}.
\end{thebibliography}
\end{document}